\documentclass{article}

\usepackage[english]{babel}

\usepackage[letterpaper,top=2cm,bottom=2cm,left=3cm,right=3cm,marginparwidth=1.75cm]{geometry}

\usepackage{amsmath,amssymb,amsfonts}
\usepackage{graphicx}
\usepackage{tabularx}
\usepackage{booktabs}
\usepackage{lscape}
\usepackage[colorlinks=true, allcolors=blue]{hyperref}

\title{Anomalous diffusion for mass transport phenomena I: Analytic solutions to time fractional diffusion}
\author{Nathaniel G. Hermann, M. Shane Hutson* \\
\\
\textit{Department of Physics, Vanderbilt University, Nashville, TN, USA} \\
\textsuperscript{*}\textit{Corresponding author: shane.hutson@vanderbilt.edu}
}
\date{}

\begin{document}

\maketitle

\begin{abstract}
     Mass transport problems are ubiquitous in diverse fields of physics and engineering. With the development of fractional calculus, many have taken to studying problems of fractional mass transport either through numerical simulations or through complex mathematical structures (\textit{e.g.} Fox-H functions). Here, we present a set of analytic solutions to common time fractional diffusion problems, written in terms of Mittag-Leffler and M-Wright functions, as well as generalized fractional error and complementary error functions derived within. We additionally show how time fractional diffusion is a special case of a two-parameter stretched-time fractional diffusion process. Finally we present a procedure to take canonical solutions to mass transport problems with Fickian diffusion and extend these to systems with anomalous diffusion.
\end{abstract}

\section*{Introduction}
Mass transport phenomena are some of the most studied problems in physics and engineering, spanning spatio-temporal ranges from the sub-cellular to the astronomical. These phenomena are typically modeled in part by Fickian diffusion, which describes the evolution of concentration $c(\mathbf{r},t)$ in a medium with diffusivity $D$ through Fick's second law,
\begin{equation}
    \frac{\partial c}{\partial t} = D\nabla^{2}c
\end{equation}
This partial differential equation (PDE), identical to the heat equation, is perhaps the most studied equation in mathematical physics, with many canonical analytic solutions in different systems \cite{Crank1975,Jaegar1986}. It may also be extended into more complicated models, such as the advection-diffusion equation, that couple mass transport with momentum transport. Systems that obey Fick's second law are said to be \textit{Fickian}, and this model accurately describes a variety of physical phenomena, such as diffusion of dilute solutes. \\
\indent However, it is known that some systems show non-Fickian behavior. Anomalous diffusion, in which the spatial spread of concentration scales nonlinearly with respect to time, arises in a variety of systems including diffusion in disordered media (\textit{e.g.} glassy media such as polymers) \cite{Muller-Plathe1992,Paul2002,Guo2012,Goychuk2018,Singh2020} or diffusion in non-dilute solutions (\textit{e.g.} intracellular protein transport) \cite{Weiss2004,Banks2005,Guigas2008,Dix2008,Sokolov2012,Sabri2020} which display subdiffusive (that is, slower than expected) behavior; and active transport, which displays superdiffusive (that is, faster than expected) behavior \cite{Bruno2009,Reverey2015}. Anomalous diffusion is characterized for a single-point concentration source by a non-linear evolution of the mean-squared displacement (MSD), $\langle x^{2}\rangle\sim D_{\nu}t^{\nu}$, where $\nu$ is the diffusive order. Subdiffusion is characterized by $0<\nu<1$, Fickian diffusion by $\nu =1$, and superdiffusion by $1<\nu<2$. A great number of models have been developed for anomalous diffusion processes \cite{Metzler2000,Metzler2014}, and this non-linear temporal scaling generally suggests the need for self-similar behavior corresponding to a power law. We will consider two models to describe such behavior: a \textit{stretched-time} (ST) diffusion equation, featuring a time-dependent diffusion constant; and a \textit{time fractional} (TF) diffusion equation, directly implementing nonlinear temporal scaling through the use of fractional calculus. \\
\indent A system with ST behavior exhibits Fickian diffusion which slows down or speeds up corresponding to a power law with respect to time. This can be modeled through the ST diffusion equation
\begin{equation}
    \frac{\partial c}{\partial(t^{\alpha})} = D_{\alpha}\nabla^{2}c
    \label{eqn:STDiffusion}
\end{equation}
This can be equivalently stated using a time-dependent diffusion constant $D_{\alpha}(t) = \alpha t^{\alpha-1}D_{\alpha}$ by using the rule
\begin{equation*}
    \frac{\partial}{\partial(t^{\alpha})} = \frac{1}{\alpha t^{\alpha-1}}\frac{\partial}{\partial t}
\end{equation*}
to obtain
\begin{equation}
    \frac{\partial c}{\partial t} = \alpha t^{\alpha-1}D_{\alpha}\nabla^{2}c
\end{equation}
Systems with this behavior have an underlying stochastic process of fractional Brownian motion (fBm) \cite{Mainardi2010,Balcerek2022,Grzesiek2024}. This process may arise from a medium with time-evolving properties, or spatially evolving properties that effectively rescale time.
\\
\indent Alternately, methods have been developed to use fractional calculus to model anomalous diffusion through a modified TF diffusion equation:
\begin{equation}
    \mathcal{D}_{t}^{\beta}c = D_{\beta}\nabla^{2}c
    \label{eqn:TFDiffusuon}
\end{equation}
Here, in place of the standard time derivative in Fick's second law, we use a Caputo fractional time derivative $\mathcal{D}_{t}^{\beta}c$). Systems with this behavior have an underlying stochastic process of gray Brownian motion (gBm) \cite{Mainardi2010,Schneider1990}. This process corresponds to a process with spatial step-sizes chosen from a non-Gaussian probability density function (PDF) arising from a heterogeneous medium with both long and short waiting times. This equation has been solved analytically for some limited systems, such as the Cauchy and signaling problems in the infinite and semi-infinite domain \cite{Mainardi2010ch6}. However, much work in TF diffusion has either used extensive numerical simulations, or been developed in complex mathematical formulations, such as the Fox-H functions \cite{Metzler2000,Mainardi2005}. In our opinion, this limits the utility of this model in the broader physics and engineering community. As fractional calculus becomes fully developed, and researchers explore not only the fractional diffusion (or heat) equation, but other fractional systems, there is a need for simple analytic models to both approximate more complex systems, and make the fractional calculus easier to manipulate. \\
\indent To this end, we have derived and cataloged various analytic solutions to the TF diffusion equation. These solutions make use of canonical fractional calculus functions -- the Mittag-Leffler function ($E$) and the M-Wright, or Mainardi, function ($M$) -- and of newly defined generalized fractional equivalents to the error function ($N$) and complementary error function ($K$). We discuss interesting properties of these new functions and then develop various solutions in various domains including: the infinite domain $\mathbb{R}$; the semi-infinite domain $\mathbb{R}^{+}$; and a finite domain. We will also discuss how to solve problems in higher dimensional systems, especially in the context of radial dependence and angular symmetry. Ultimately, we will show how both ST and TF diffusion are special cases of generalized \textit{stretched-time fractional} (STF) diffusion, and how one may easily translate canonical solutions to the Fickian diffusion equation to any of these three anomalous diffusion equations.
\section*{A Review of Fractional Calculus}
\begin{figure}
    \centering
    \includegraphics[scale = 0.6]{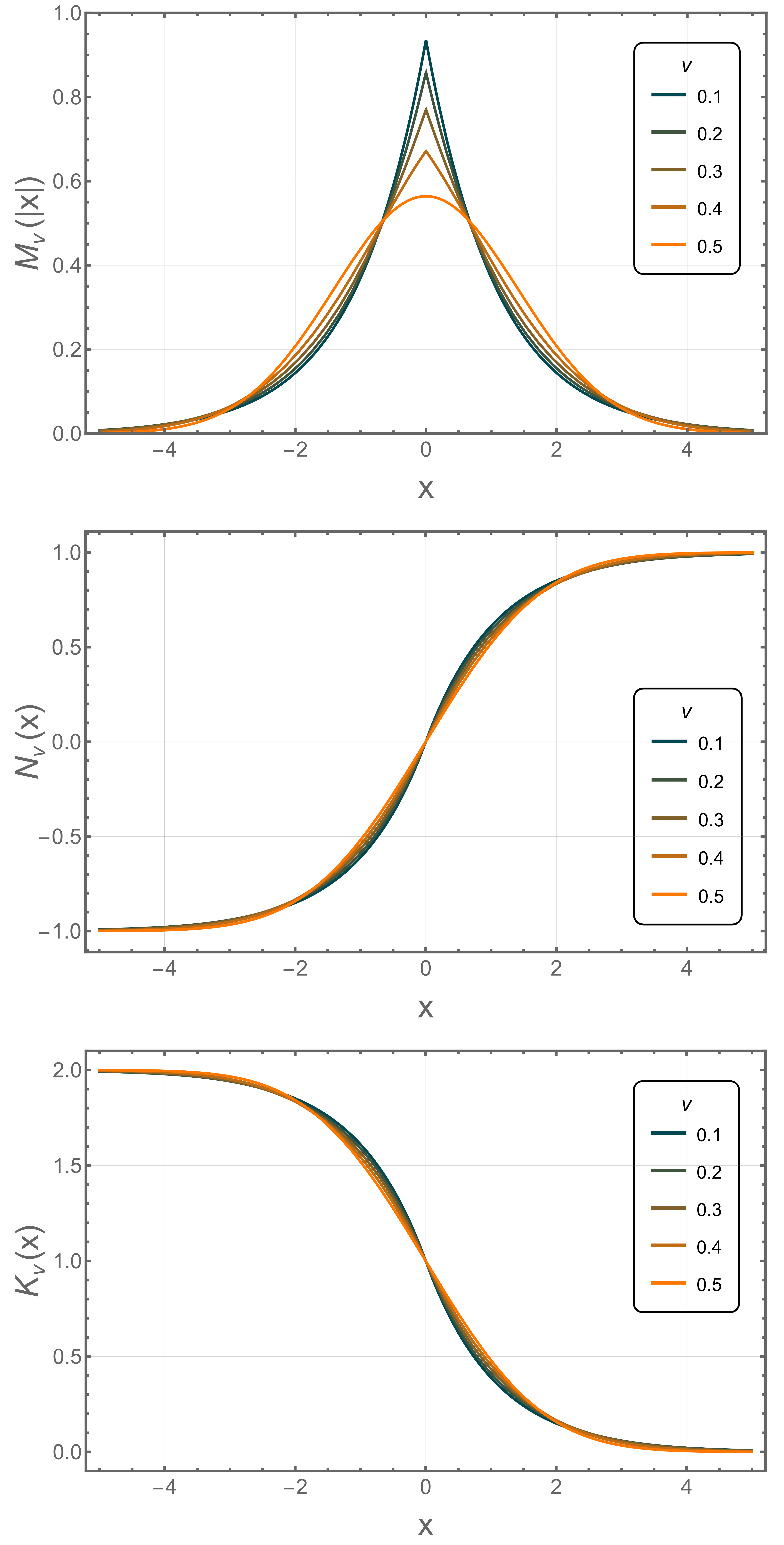}
    \caption{Plots of the M-Wright (top), fractional error (middle), and fractional complementary error functions (bottom) over TF derivative order $\nu$, where $\nu = 0.5$ recovers Fickian diffusion.}
    \label{fig:functionPlots}
\end{figure}

Fractional calculus is, in essence, the extension of integrals and derivatives to non-integer orders. Consider the fractional operator known as the Riemann-Liouville integral $_{RL}\mathcal{I}_{t}^{\mu}$. This is defined for any positive, real $\mu$ through the convolution integral
\begin{equation*}
    _{RL}\mathcal{I}_{t}^{\mu}f(t) = \frac{1}{\Gamma(\mu)}\int_{0}^{t}(t-\tau)^{\mu-1}f(\tau)d\tau
\end{equation*}
Note that $_{RL}\mathcal{I}_{t}^{\mu}\ _{RL}\mathcal{I}_{t}^{1-\mu} = \mathbb{I}$ where $\mathbb{I}$ is the identity operator, and that 
\begin{equation*}
_{RL}\mathcal{I}_{t}^{\mu}\ _{RL}\mathcal{I}_{t}^{\nu} = _{RL}\mathcal{I}_{t}^{\nu}\ _{RL}\mathcal{I}_{t}^{\mu} = _{RL}\mathcal{I}_{t}^{\mu+\nu}
\end{equation*}
Similarly, one can define the Riemann-Liouville fractional derivative $_{RL}\mathcal{D}_{t}^{\mu}$ as the left-inverse of the Riemann-Liouville integral such that $_{RL}\mathcal{D}_{t}^{\mu}\ _{RL}\mathcal{I}_{t}^{\mu} = \mathbb{I}$. It follows then that for $m\in\mathbb{N}$ and $m-1<\mu<m$, the derivative is defined through the convolution integral
\begin{equation*}
    _{RL}\mathcal{D}_{t}^{\mu}f(t) = \frac{d^{m}}{dt^{m}}\left[\frac{1}{\Gamma(m-\mu)}\int_{0}^{t}\frac{f(\tau)}{(t-\tau)^{\mu+1-m}}d\tau\right]
\end{equation*}
\indent From these definitions, we can define the Caputo fractional derivative $_{C}\mathcal{D}_{t}^{\nu}$ such that $_{C}\mathcal{D}_{t}^{\nu} = _{RL}\mathcal{I}_{t}^{\mu-\nu}\ _{RL}\mathcal{D}_{t}^{\mu}$. This derivative has convolution integral definition
\begin{equation*}
     _{C}\mathcal{D}_{t}^{\nu} f(t)= \frac{1}{\Gamma(1-\nu)}\int_{0}^{t}(t-\tau)^{-\nu}\left(\frac{\partial f}{\partial \tau}\right)d\tau
\end{equation*}
The Riemann-Liouville and Caputo fractional derivatives are related:
\begin{equation*}
    _{C}\mathcal{D}_{t}^{\nu}[f(t)] = _{RL}\mathcal{D}_{t}^{\nu}[f(t)] - \sum_{n=0}^{\lceil\nu\rceil}\frac{t^{n-\nu}}{\Gamma(n-\nu-1)}f^{(n)}(0)
\end{equation*}
This is revealing in so far as the Caputo derivative is, in a sense, a regularization of the Riemann-Liouville fractional derivative. Herein, to describe TF diffusion, we will use a time fractional derivative of the Caputo form and drop the subscript $_{C}\mathcal{D}_{t}^{\nu} = \mathcal{D}_{t}^{\nu}$. Usefully, the Caputo fractional derivative has well-defined Laplace transform $\mathcal{L}:t\rightarrow s$,
\begin{equation}
    \mathcal{L}\left[\mathcal{D}_{t}^{\nu}f(t)\right] = s^{\nu}F(s) - \sum_{n=0}^{\lceil\nu\rceil}s^{\nu-n-1}f^{(n)}(0)
\end{equation}
Note that integral transforms used throughout are defined in Appendix A. \\
\indent Solutions to fractional calculus problems often involve two special functions: the Mittag-Leffler function, $E$, and the M-Wright, or Mainardi, function $M$. The one-parameter Mittag-Leffler function was first introduced by Gösta Mittag-Leffler in 1903 \cite{Mittag-Leffler1903}, and this well-studied function arises in many fractional calculus contexts \cite{Haubold2011,Mainardi2010aE,Mainardi2020}. The series definition of the Mittag-Leffler function, with order $\nu$, is
\begin{equation}
    E_{\nu}(z) = \sum_{n=0}^{\infty}\frac{z^{n}}{\Gamma(\nu n+1)}
\end{equation}
Many special cases of the Mittag-Leffler function have been tabulated, including 
\begin{equation}
    E_{1}(z) = \exp(z)
\end{equation}
and readers interested in further properties of this function may consult the thorough review of Haubold et al. \cite{Haubold2011}. Numerical implementations of the Mittag-Leffler function are available in many common programming languages, including Python, R, MATLAB, and Mathematica. \\
\indent The M-Wright function is a auxillary function, first described by Francesco Mainardi in 1994 \cite{Mainardi1994}, related to the generalized Wright function $W_{\lambda,\mu}$ \cite{Wright1933}, defined as
\begin{equation}
    M_{\nu}(z) = W_{-\nu,1-\nu}(-z)
\end{equation}
This function has been well studied as a fundamental solution to problems in fractional calculus \cite{Mainardi2010,Pagnini2013}. The series definition of the M-Wright function is then given as
\begin{equation}
    M_{\nu}(z) = \sum_{n=0}^{\infty}\frac{(-z)^{n}}{n!\Gamma(-\nu n + (1-\nu))} = \frac{1}{\pi}\sum_{n=1}^{\infty}\frac{(-z)^{n-1}}{(n-1)!}\Gamma(\nu n)\sin{(\pi\nu n)}
\end{equation}
while the integral definition over the Hankel loop ($Ha$) is
\begin{equation}
    M_{\nu}(z) = \frac{1}{2\pi i}\int_{Ha}\exp{(\sigma-z\sigma^{\nu})}\frac{d\sigma}{\sigma^{1-\nu}}
\end{equation}
Plots of the M-Wright function for a subdiffusive system are shown in Figure \ref{fig:functionPlots}. Many special cases of the M-Wright function have been tabulated, including
\begin{equation}
    M_{1/2}(z) = \frac{1}{\sqrt{\pi}}\exp{\left(-\frac{z^{2}}{4}\right)}
\end{equation}
and important integral transforms of these functions for values of $0<\nu<1$ have been explicitly calculated. These include Laplace transform,
\begin{equation}
    \mathcal{L}[M_{\nu}(t)] = E_{\nu}(-s)
\end{equation}
the Fourier transform of the symmetrized M-Wright function,
\begin{equation}
    \mathcal{F}[M_{\nu}(|x|)] = 2E_{2\nu}(-k^{2})
    \label{eqn:MWrightFourierTrans}
\end{equation}
the Mellin transform,
\begin{equation}
    \mathcal{M}[M_{\nu}(r)] = \frac{\Gamma(s)}{\Gamma(\nu(s-1)+1)}
\end{equation}
as well as the exponential Laplace transform pairs,
\begin{align}
    \mathcal{L}\left[\frac{\nu}{t^{\nu+1}}M_{\nu}\left(\frac{1}{t^{\nu}}\right)\right] &= \exp{(-s^{-\nu})} \label{eqn:MWrightExpTrans1} \\
    \mathcal{L}\left[\frac{1}{t^{\nu}}M_{\nu}\left(\frac{1}{t^{\nu}}\right)\right] &= \frac{\exp{(-s^{-\nu})}}{s^{1-\nu}}
\end{align}
The absolute moments in $\mathbb{R}^{+}$ can be directly calculated for $0\leq \nu<1$
\begin{equation}
    \int_{0}^{\infty}x^{\delta}M_{\nu}(x)dx = \frac{\Gamma(\delta+1)}{\Gamma(\nu\delta+1)}
\end{equation}
As the symmetrized M-Wright function when normalized is a PDF, the above result can directly determine the moments of the PDF. Readers interested further in this function should refer to the appendix of Mainardi's monograph on fractional calculus \cite{Mainardi2010aF}. Currently, there are no commonly used programming languages with well-supported numerical approximations of the M-Wright function. 

\section*{Fractional Error Functions}
To develop analytic solutions to some fractional diffusion problems, it is necessary introduce two new special functions. Consider the relation between the Gaussian and the error function, $\text{erf}(x)$,
\begin{equation*}
    \text{erf}(z) = \frac{2}{\sqrt{\pi}}\int_{0}^{z}\exp{(-\xi^{2})}d\xi
\end{equation*}
As discussed, the symmetrized M-Wright function when normalized in a PDF and simplifies to a Gaussian explicitly at $\nu = 1/2$. Therefore, one can construct a fractional error function, $N_{\nu}(z)$, analogous to the error function,
\begin{equation}
    N_{\nu}(z) = \int_{0}^{z}M_{\nu}(|\xi|)d\xi
\end{equation}
The series form of this fractional error function is,
\begin{equation}
    N_{\nu}(z) = \sum_{n=0}^{\infty}\frac{(-1)^{n}\text{sgn}(z)|z|^{n+1}}{(n+1)!\Gamma(-\nu n + (1-\nu))}
\end{equation}
Then, we can introduce a second function: the fractional complementary error function, $K_{\nu}(z)$ which plays an analogous role to the normal complementary error function,
\begin{equation}
    K_{\nu}(z) = \int_{z}^{\infty}M_{\nu}(|\xi|)d\xi = 1 - N_{\nu}(x)
\end{equation}
Plots of $N_{\nu}(x)$ and $K_{\nu}(x)$ are shown in Figure \ref{fig:functionPlots}. When considering the case that $\nu = 1/2$, we recover the standard error functions,
\begin{align}
    N_{1/2}(z) &= \text{erf}\left(\frac{z}{2}\right) \\
    K_{1/2}(z) &= \text{erfc}\left(\frac{z}{2}\right)
\end{align} 
\indent Consider now the integral transforms of these new functions for $0<\nu<1$. The Laplace transforms are
\begin{align}
    \mathcal{L}[N_{\nu}(t)] &= \frac{1}{s}E_{\nu}(-s) \label{eqn:NLap}\\
    \mathcal{L}[K_{\nu}(t)] &= \frac{1}{s}(1-E_{\nu}(-s)) \label{eqn:KLap}
\end{align}
The Fourier transforms are
\begin{align}
    \mathcal{F}[N_{\nu}(x)] &= \frac{2}{ik}E_{2\nu}(-k^{2}) \\
    \mathcal{F}[K_{\nu}(x)] &= \sqrt{2\pi}\delta(k) - \frac{2}{ik}E_{2\nu}(-k^{2})
\end{align}
The Mellin transforms are
\begin{align}
    \mathcal{M}[N_{\nu}(r)] &= -\frac{\Gamma(s+1)}{s\Gamma(\nu s+1)} \\ 
    \mathcal{M}[K_{\nu}(r)] &= +\frac{\Gamma(s+1)}{s\Gamma(\nu s+1)}
\end{align}
Note that the definitions of the Laplace transforms of the fractional error functions (Eqns. \ref{eqn:NLap}-\ref{eqn:KLap}) are similar but distinct to the canonical Laplace transform of the Wright function for $-1\leq\nu< 0$ \cite{Mainardi2020a},
\begin{equation*}
    \mathcal{L}[W_{\nu,\mu+\nu}(-t)] = E_{-\nu,\mu-\nu}(-s)
\end{equation*}
Indeed, these fractional error functions cannot be expressed as a generalized Wright function.

\section*{Translating Canonical Solutions to Stretched-Time Systems}
It is the case that many analytic and semi-analytic solutions have been derived for the diffusion equation (or equivalently, for the heat equation) \cite{Crank1975,Jaegar1986}. Rather than solve these problems in their ST form from first principles, we can see that one can translate these solutions from their Fickian solution into the equivalent time-stretched solution with ease. Recognizing that Eqn. \ref{eqn:STDiffusion} is identical to the standard diffusion equation with time variable $\tau = t^{\alpha}$, we can quickly derive any time-stretched solution by making the substitution $Dt\rightarrow D_{\alpha}t^{\alpha}$. For instance, the 1D ST diffusion equation in $\mathbb{R}$ has Green's function \cite{Mainardi2010}
\begin{equation}
    G(x,t) = \frac{1}{\sqrt{4\pi D_{\alpha}t^{\alpha}}}\exp{\left(-\frac{x^{2}}{4D_{\alpha}t^{\alpha}}\right)}
\end{equation}
When solving problems in a system with TF diffusion, keep in mind that these systems all have simple solutions in the ST formulation.

\section*{Analytic Solutions to the 1D Time Fractional Diffusion Equation}

\subsection*{Solutions in Infinite Media}
First, let us consider one dimensional TF problems where $x\in\mathbb{R}$ with diffusivity $D_{\beta}$. We shall restrict $0<\beta<1$ to obtain the subdiffusive solutions. Systems with superdiffusion are discussed in Appendix B. The same problem in a system with space-fractional diffusion is discussed in Appendix C.
\subsubsection*{Plane Source}

\begin{figure}
    \centering
    \includegraphics[scale = 1.25]{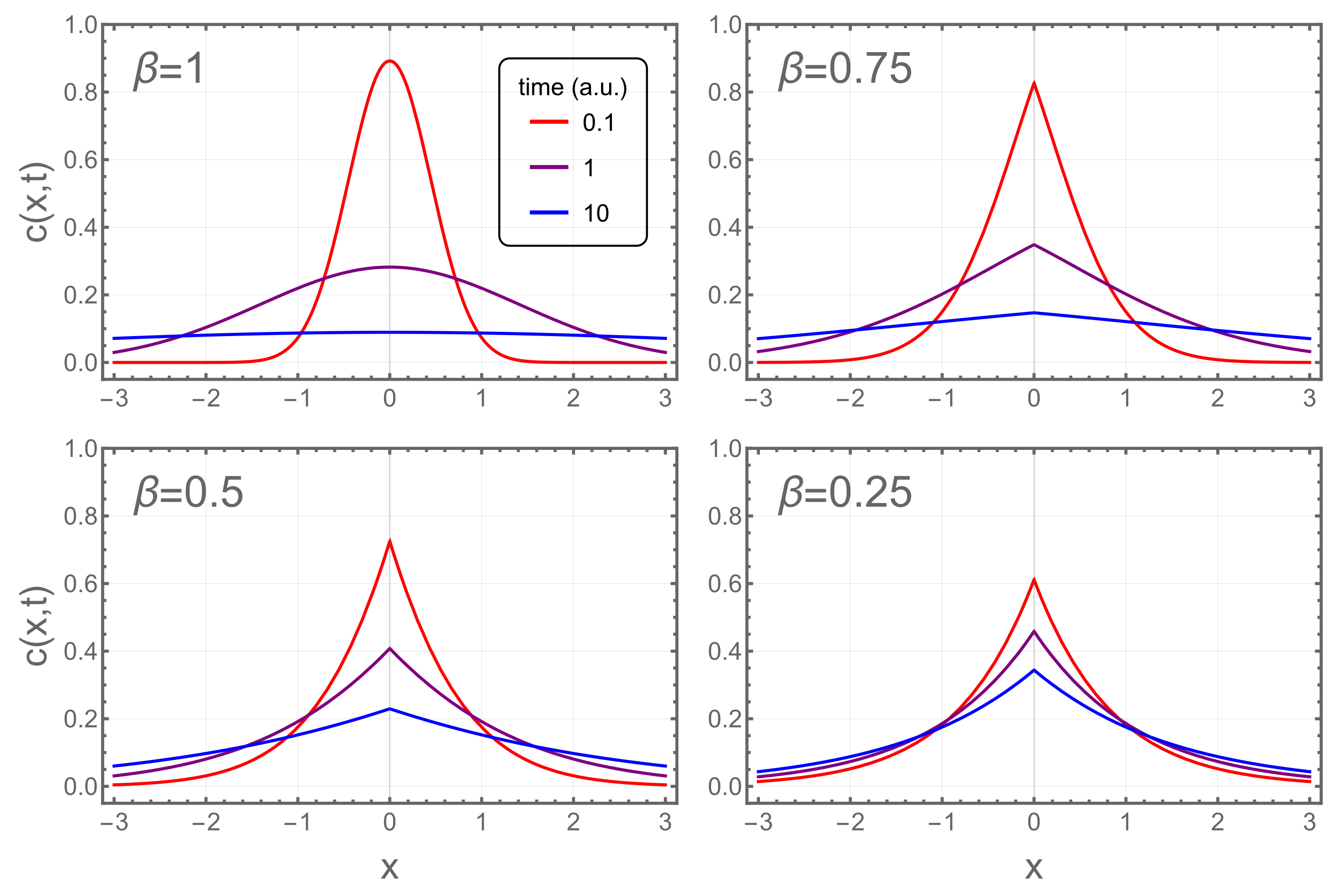}
    \caption{TF diffusion in an infinite domain with initial condition $c(x,t) = N_{tot}\delta(x)$. Plotted results of (Eqn. \ref{eqn:DeltaICInfiniteSol}) with $N_{tot} = 1$ and $D_{\beta}=1$.}
    \label{fig:InfDomDeltaIC}
\end{figure}

\indent Suppose we have initial condition $c(t=0) = N_{tot}\delta(x)$. We may apply the Laplace transform $\mathcal{L}[c(x,t)] = \tilde{c}(x,s)$ to the 1D TF diffusion PDE to obtain
\begin{equation*}
    s^{\beta}\tilde{c}(x,s) - s^{\beta-1}N_{tot}\delta(x) = D_{\beta}\frac{\partial^{2}\tilde{c}}{\partial x^{2}}
\end{equation*}
We then apply a Fourier transform to this PDE and recover
\begin{equation*}
    s^{\beta}\hat{\tilde{c}}(k,s) - s^{\beta-1}N_{tot} = -D_{\beta}k^{2}\hat{\tilde{c}}(k,s)
\end{equation*}
It is generally known \cite{Mainardi2010aE} that 
\begin{equation*}
    \mathcal{L}[E_{\beta}(-\lambda t^{\beta})] = \frac{s^{\beta-1}}{s^{\beta} + \lambda}
\end{equation*}
and so, taking the inverse Laplace transform, one finds
\begin{equation*}
    \hat{c}(k,t) = N_{tot}E_{\beta}(-D_{\beta}k^{2}t^{\beta})
\end{equation*}
Finally, given the definition of the Fourier transform of the M-Wright function (Eqn. \ref{eqn:MWrightFourierTrans}), one recovers the solution given by the inverse Fourier transform,
\begin{equation}
    c(x,t) = \frac{N_{tot}}{\sqrt{4D_{\beta}t^{\beta}}}M_{\beta/2}\left(\frac{|x|}{\sqrt{D_{\beta}t^{\beta}}}\right)
    \label{eqn:DeltaICInfiniteSol}
\end{equation}
Similar Fourier-Laplace transform approaches are indeed appropriate for all initial value problems in the infinite 1D Cartesian domain. This solution satisfies all requirements, \textit{i.e.} $c(x\rightarrow\pm\infty) = 0$ and
\begin{equation*}
    \int_{-\infty}^{\infty}c(x,t)dx = N_{tot}
\end{equation*}
When $\beta = 1$, \textit{i.e.} the Fickian case, we recover the standard form of the Fickian solution,
\begin{equation*}
    c(x,t) = \frac{N_{tot}}{\sqrt{4\pi Dt}}\exp\left(-\frac{x^{2}}{4Dt}\right)
\end{equation*}
This solution recalls the Green's function for TF diffusion on the infinite domain, scaled by $N_{tot}$ \cite{Mainardi2010},
\begin{equation}
    G_{\beta}(x,t) = \frac{1}{\sqrt{4D_{\beta}t^{\beta}}}M_{\beta/2}\left(\frac{|x|}{\sqrt{D_{\beta}t^{\beta}}}\right)
    \label{eqn:1DTFInfiniteGreensFunc}
\end{equation}
Plots of (Eqn. \ref{eqn:DeltaICInfiniteSol}) are shown in Figure \ref{fig:InfDomDeltaIC} for varying $\beta$.

\subsubsection*{Extended Sources}
\begin{figure}
    \centering
    \includegraphics[scale = 1.25]{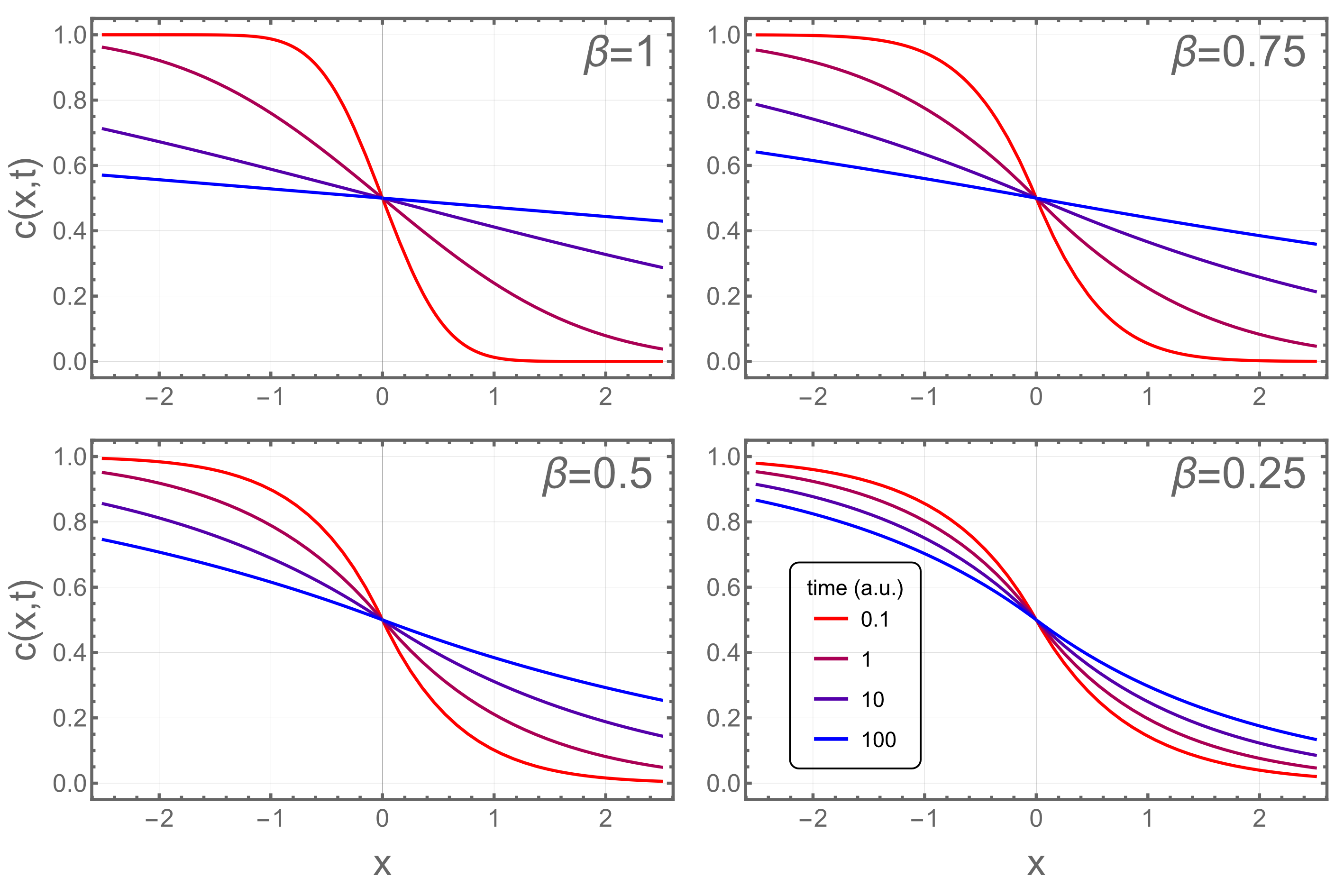}
    \caption{TF diffusion in an infinite domain with initial condition $c(x,t) = c_{0}\Theta(-x)$. Plotted results of (Eqn. \ref{eqn:HeavisideICInfiniteSol}) with $c_{0} = 1$ and $D_{\beta}=1$.}
    \label{fig:InfDomHeavisideIC}
\end{figure}

Consider the system with initial condition $c(t=0) = c_{0}\Theta(-x)$, where $\Theta$ is the Heaviside function. The Laplace transformed PDE is:
\begin{equation*}
    s^{\beta}\tilde{c}(x,s) - s^{\beta-1}c_{0}\Theta(-x) = D_{\beta}\frac{\partial^{2}\tilde{c}}{\partial x^{2}}
\end{equation*}
The subsequent equation in Fourier-Laplace space is
\begin{equation*}
    s^{\beta}\hat{\tilde{c}}(k,s) +\left(\frac{ic_{0}}{k} - c_{0}\pi\delta(k)\right)s^{\beta-1} = -D_{\beta}k^{2}\hat{\tilde{c}}(k,s)
\end{equation*}
Rearranging,
\begin{equation*}
    \hat{\tilde{c}}(k,s) = -\frac{ic_{0}}{k}\frac{s^{\beta-1}}{s^{\beta}+D_{\beta}k^{2}} + c_{0}\pi\delta(k)\frac{s^{\beta-1}}{s^{\beta}+D_{\beta}k^{2}}
\end{equation*}
Using the inverse transforms of the fractional complementary error function, the solution to this problem is,
\begin{equation}
    c(x,t) = \frac{c_{0}}{2}K_{\beta/2}\left(\frac{x}{\sqrt{D_{\beta}t^{\beta}}}\right)
    \label{eqn:HeavisideICInfiniteSol}
\end{equation}
Plots of (Eqn. \ref{eqn:HeavisideICInfiniteSol}) at various $\beta$ are shown in Figure \ref{fig:InfDomHeavisideIC}. Again, we can observe that for the Fickian case when $\beta = 1$, we recover the canonical solution
\begin{equation*}
    c(x,t) = \frac{c_{0}}{2}\text{erfc}\left(\frac{x}{\sqrt{4Dt}}\right)
\end{equation*}

\subsection*{Solutions in Semi-Infinite Media}
Now, consider the one dimensional TF problem in which $x\in\mathbb{R}^{+}$ with diffusivity $D_{\beta}$. We shall restrict $0<\beta<1$ to obtain the subdiffusive solutions.
\subsubsection*{Constant Surface Concentration}
\begin{figure}
    \centering
    \includegraphics[scale = 1.25]{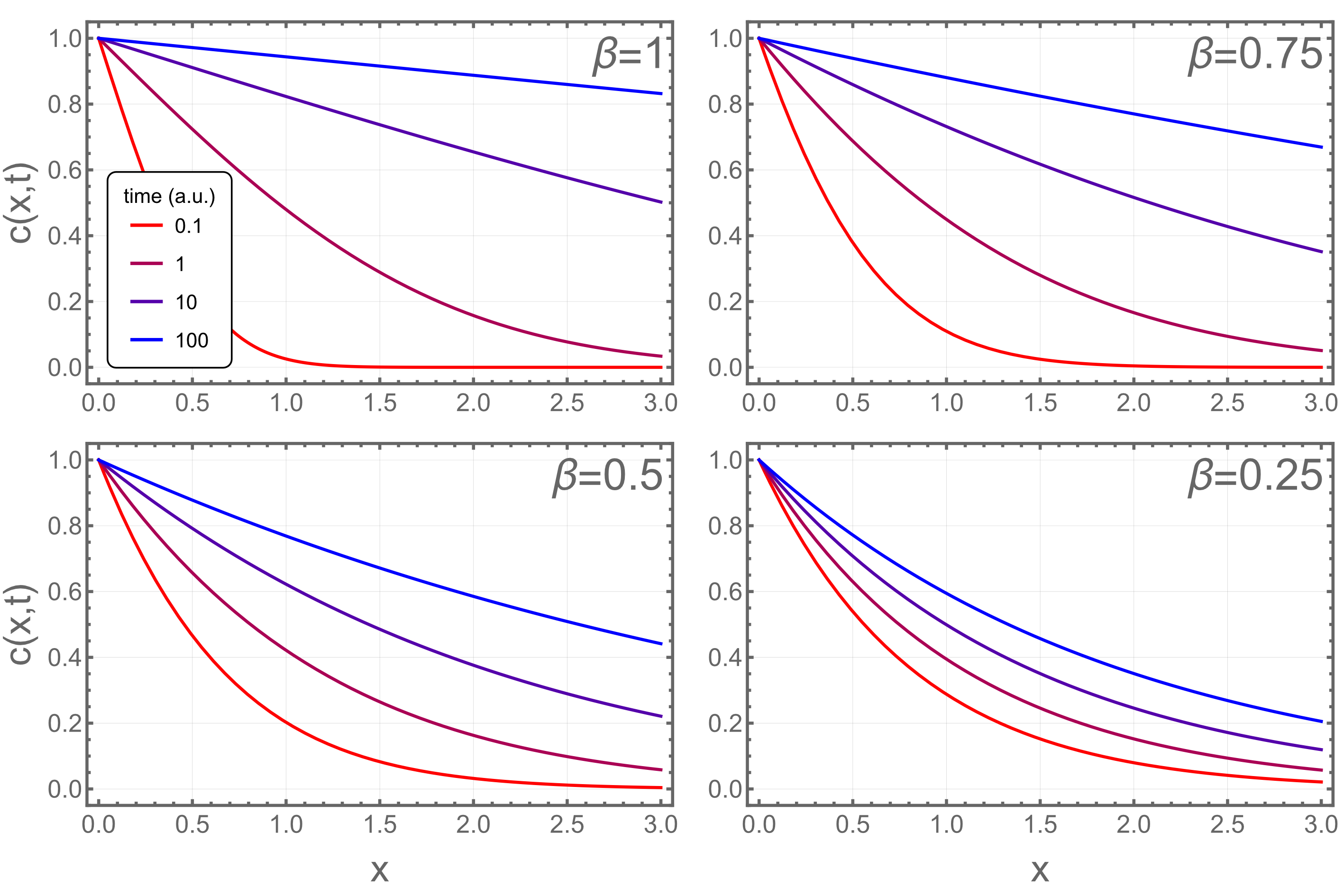}
    \caption{Fractional diffusion in an semi-infinite domain ($x\geq0$) with initial condition $c(x>0,t) = 0$ and boundary condition $c(x=0,t) = c_{0}$. Plotted results of (Eqn. \ref{eqn:SemiInfFixedBCSol}) with $c_{0} = 1$ and $D_{\beta}=1$.}
    \label{fig:SemiInfFixedBC}
\end{figure}

Consider the system with an initial concentration $c(x>0,t=0) = 0$ and a fixed boundary concentration $c(x=0) = c_{0}$. This problem is described by Mainardi as the signaling problem \cite{Mainardi2010ch6}, and for a boundary value of $c(x=0) = h(t) = c_{0}$, this problem has Green's function
\begin{equation*}
    G(x,t) = \frac{\beta x}{\sqrt{4D_{\beta}t^{3\beta}}}M_{\beta/2}\left(\frac{x}{\sqrt{D_{\beta}t^{\beta}}}\right)
\end{equation*}
The general solution to this problem can be developed through the convolution of this Green's function with the boundary value, $G*h$. Taking the Laplace transform of both functions,
\begin{equation*}
    \tilde{G}(x,s) = \mathcal{L}\left[\frac{(\beta/2) x}{\sqrt{D_{\beta}}t^{3\beta/2}}M_{\beta/2}\left(\frac{x}{\sqrt{D_{\beta}}t^{\beta/2}}\right)\right]
\end{equation*}
Using the Laplace transform (Eqn. \ref{eqn:MWrightExpTrans1}) and standard Laplace transform scaling rules, we find
\begin{equation*}
    \tilde{G}(x,s) = \frac{\sqrt{D_{\beta}}}{x}\exp{\left(-\frac{\sqrt{D_{\beta}}}{x}s^{-\beta/2}\right)}
\end{equation*}
Along with $\tilde{h}(s) = s^{-1}$, the general solution is then
\begin{equation}
    c(x,t) = \mathcal{L}^{-1}\left[\frac{1}{s}\frac{\sqrt{D_{\beta}}}{x}\exp{\left(-\frac{\sqrt{D_{\beta}}}{x}s^{-\beta/2}\right)}\right]
\end{equation}
Now, recall that for Laplace transform pair $f(t)$ and $\tilde{f}(s)$, $\mathcal{L}^{-1}[s^{-1}\tilde{f}(s)] = \int_{0}^{t}f(\tau)d\tau$. With this, it follows that
\begin{equation*}
    c(x,t) = \int_{0}^{t}\frac{(\beta/2) x}{\sqrt{D_{\beta}}\tau^{3\beta/2}}M_{\beta/2}\left(\frac{x}{\sqrt{D_{\beta}}\tau^{\beta/2}}\right)d\tau
\end{equation*}
Through a substitution ($\xi = x/\sqrt{D_{\beta}\tau^{\beta}}$) it follows that
\begin{equation}
    c(x,t) = -\int_{\infty}^{x/\sqrt{D_{\beta}t^{\beta}}}M_{\beta/2}(\xi)d\xi = \int^{\infty}_{x/\sqrt{D_{\beta}t^{\beta}}}M_{\beta/2}(\xi)d\xi
\end{equation}
And so, the solution is in terms of the fractional complementary error function $K$,
\begin{equation}
    c(x,t) = c_{0}K_{\beta/2}\left(\frac{x}{\sqrt{D_{\beta}t^{\beta}}}\right)
    \label{eqn:SemiInfFixedBCSol}
\end{equation}
Note that when $\beta=1$, this reduces to the standard Fickian solution,
\begin{equation*}
    c(x,t) = c_{0}\text{erfc}\left(\frac{x}{\sqrt{4Dt}}\right)
\end{equation*}
The methods used above are appropriate for any problem in the semi-infinite positive real line. Plots of (Eqn. \ref{eqn:SemiInfFixedBCSol}) for varying $\beta$ are shown in Figure \ref{fig:SemiInfFixedBC}.

\subsection*{Solutions in Finite Plane Sheet}
Now, let us consider one dimensional TF problems where $x$ is restricted to an interval in $\mathbb{R}$, and with diffusivity $D_{\beta}$. We shall restrict $0<\beta<1$ to obtain the subdiffusive solutions.
\subsubsection*{Equal, Fixed Boundary Concentrations}
\begin{figure}
    \centering
    \includegraphics[scale = 1.25]{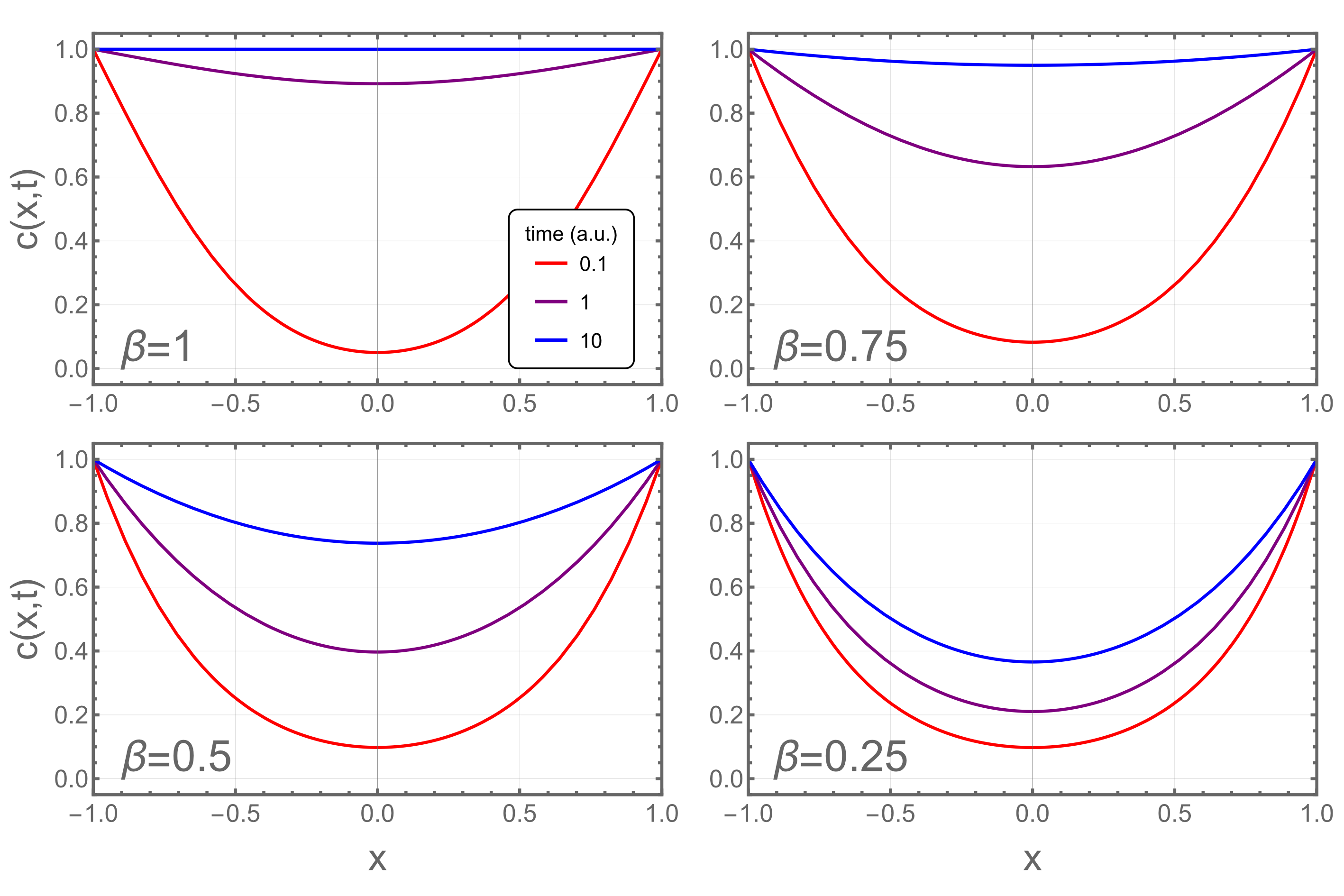}
    \caption{Fractional diffusion in an semi-infinite domain ($x\geq0$) with initial condition $c(x>0,t) = 0$ and boundary condition $c(x=0,t) = c_{0}$. Plotted results of (Eqn. \ref{eqn:FiniteEqualBCSol}) with $c_{0} = 1$ and $D_{\beta}=1$.}
    \label{fig:FiniteFixedEqualBC}
\end{figure}

Consider a system in which $x\in[-L,L]$ with fixed boundary values $c(x=\pm L) = c_{0}$ and initial concentration $c(|x|<L,t=0) = 0$. This problem may be solved through the separation of variables method, letting $c(x,t) = X(x)T(t)$. This produces the separated equations
\begin{align*}
    \mathcal{D}_{t}^{\nu}T &= -D\lambda^{2} T \\
    \frac{d^{2}X}{dx^{2}} &= -\lambda^{2} X
\end{align*}
The latter spatial ODE produces the typical trigonometric series, but the former fractional temporal ODE is in fact the fractional relaxation equation, which has a known result in terms of the Mittag-Leffler function,
\begin{equation*}
    T(t)\sim E_{\nu}(-\lambda^{2} D_{\nu}t^{\nu})
\end{equation*}
The full solution when solved using the boundary values and initial condition is then
\begin{equation}
    c(x,t) = c_{0} - \frac{4c_{0}}{\sqrt{\pi}}\sum_{n=0}^{\infty}\frac{(-1)^{n}}{2n+1}E_{\beta}\left(-\frac{(2n+1)^{2}\pi^{2}D_{\beta}t^{\beta}}{(2L)^{2}}\right)\cos{\left(\frac{(2n+1)\pi x}{2L}\right)}
    \label{eqn:FiniteEqualBCSol}
\end{equation}
Plots of (Eqn. \ref{eqn:FiniteEqualBCSol}) with varying $\beta$ are shown in Figure \ref{fig:FiniteFixedEqualBC}.

\subsubsection*{Non-equal, Fixed Boundary Concentrations}
\begin{figure}
    \centering
    \includegraphics[scale = 1.25]{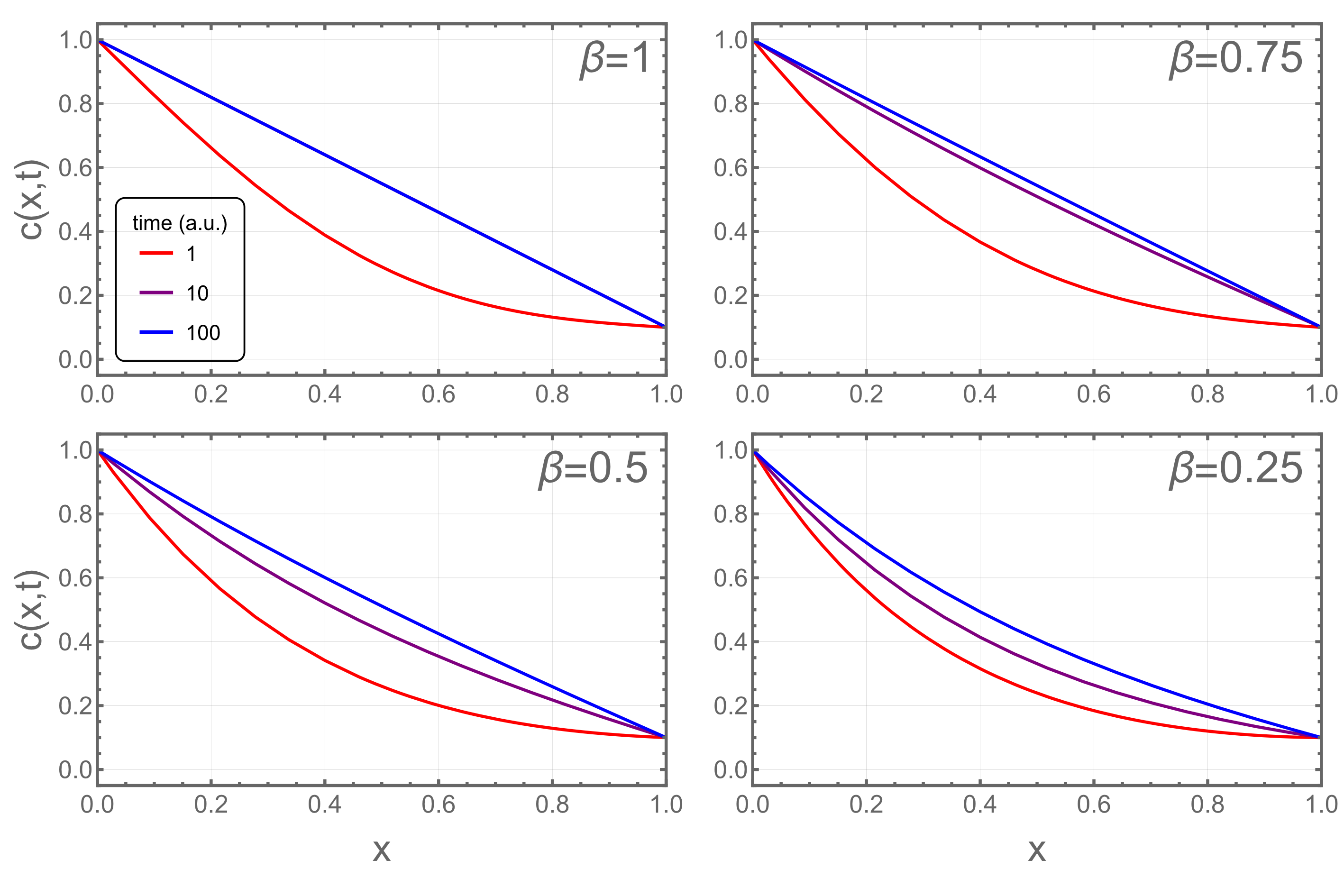}
    \caption{Fractional diffusion in an finite domain ($0\leq x\leq L$) with initial condition $c(0<x<L,t=0) = 0$ and boundary conditions $c(x=0,t) = c_{1}$ and $c(x=L,t) = c_{2}$. Plotted results of (Eqn. \ref{eqn:FiniteNonEqualBCSol}) with $L=1$, $c_{1} = 1$, $c_{2} = 0.1$, and $D_{\beta}=0.1$.}
    \label{fig:FiniteFixedNonequalBC}
\end{figure}

Now consider the system in which $x\in[0,L]$ with boundary conditions $c(x=0) = c_{1}$ and $c(x=L) = c_{2}$ and initial condition $c(0<x<L,t=0) = 0$. The solution may again be found via separation of variables, and thus the TF behavior is confined to the behavior of the time-evolution,
\begin{equation}
    c(x,t) = c_{1} + (c_{2}-c_{1})\frac{x}{L} + \frac{2}{\pi}\sum_{n=1}^{\infty}\frac{c_{2}\cos{(n\pi)}-c_{1}}{n}\sin{\left(\frac{n\pi x}{L}\right)}E_{\beta}\left(-\frac{D_{\beta}n^{2}\pi^{2}t^{\beta}}{L^{2}}\right)
    \label{eqn:FiniteNonEqualBCSol}
\end{equation}
Plots of (Eqn. \ref{eqn:FiniteNonEqualBCSol}) with varying $\beta$ are shown in Figure \ref{fig:FiniteFixedNonequalBC}.

\subsection*{Solutions in Higher Dimensions}
Thus far we have dealt only with problems in one dimension. If we instead consider an n-dimensional Euclidean space $\mathbf{r}\in\mathbb{R}^{n}$, the problem no longer permits closed solutions for most values of $\beta$. For example, the symmetric 2D TF diffusion equation on the infinite plane can be written in polar coordinates as
\begin{equation}
    \mathcal{D}_{t}^{\beta}c = D_{\beta}\left(\frac{\partial^{2}c}{\partial r^{2}} + \frac{1}{r}\frac{\partial c}{\partial r}\right)
\end{equation}
where $r\in[0,\infty)$ and with initial condition $c(r,0) = \delta(r)/r$. To find the solution to this Cauchy problem, we apply in sequence a Laplace transform and a zeroth-order Hankel transform to recover
\begin{equation*}
    s^{\beta}\tilde{C}(k,s) - s^{\beta-1} = -k^{2}D_{\beta}\tilde{C}
\end{equation*}
Taking the inverse transforms, this results in a non-closed, integral solution consisting of the inverse zeroth-order Hankel transform,
\begin{equation}
    c(r,t) = \int_{0}^{\infty}kE_{\beta}(-D_{\beta}k^{2}t^{\beta})J_{0}(kr)dk
\end{equation}
Solutions of this form have been extensively studied on the real (half-)plane \cite{Povstenko2008,Povstenko2010,Povstenko2022}. However, it is the case that no closed form solutions have been found for the infinite real space, requiring numerical methods to evaluate inverse Hankel transform integrals. Such systems highly relevant: for instance, the ubiquitous biological assay of fluorescence recovery after photobleaching (FRAP), used to determine the diffusivity of molecules in a system \cite{Axelrod1976}, will require solutions to such polar/cylindrical fractional diffusion equations to integrate this model of anomalous diffusion. \\
\indent On the other hand, on a sufficiently well-behaved subdomain of $\mathbb{R}^{n}$ such that we may develop separable solutions, we may develop analytic solutions to TF diffusion problems as for the finite plane sheet through separation of variables. In general, these solutions can be generated via replacement of the exponential temporal propagator with the Mittag-Leffler-type propagator as done herein, regardless of the form of the spatial eigenfunctions of the system (\textit{e.g.} Bessel functions, spherical harmonics).

\section*{Translating Canonical Solutions to Fractional Problems}
As we did for solutions to the ST diffusion equation, one may translate canonical solutions to the Fickian diffusion equation to solutions to the TF diffusion equation. Through the solutions developed above, one can see that one may translate these solutions into a fractional form and remain confident of their validity for systems with time fractional diffusion. In summary: for one dimensional problems on the infinite or semi-infinite Cartesian domain, Gaussian or error function solutions can be replaced as follows:
\begin{align}
    \exp{\left(-\frac{x^{2}}{4Dt}\right)} & \xrightarrow{}M_{\beta/2}\left(\frac{|x|}{\sqrt{D_{\beta}t^{\beta}}}\right) \\
    \text{erf}{\left(\frac{x}{\sqrt{4Dt}}\right)} & \xrightarrow{}N_{\beta/2}\left(\frac{x}{\sqrt{D_{\beta}t^{\beta}}}\right) \\
    \text{erfc}{\left(\frac{x}{\sqrt{4Dt}}\right)} & \xrightarrow{}K_{\beta/2}\left(\frac{x}{\sqrt{D_{\beta}t^{\beta}}}\right)
\end{align}
while solutions in finite domains require the time evolution eigenfunctions of the solution to be replaced:
\begin{equation}
    \exp{\left(-\lambda^{2}Dt\right)} \xrightarrow{} E_{\beta}\left(-\lambda^{2}D_{\beta}t^{\beta}\right)
    \label{eqn:TFMittagLefflerSub}
\end{equation}
Generally, pre-factors of $Dt\rightarrow D_{\beta}t^{\beta}$.  Given the literature that has a huge number of analytic solutions to diffusion (or heat) equations, one may reasonably select the analogous problem and translate it to a time fractional problem thusly. One must also consider modified flux boundary conditions at interfaces between domains with different diffusive orders. To match flux at such a boundary, one must consider the generalized flux defined by a Riemann-Liouville derivative or integral \cite{Povstenko2015}:
\begin{equation}
    \mathbf{J} = -D_{\beta}\ _{RL}\mathcal{D}_{t}^{1-\beta}\nabla c = -D_{\beta}\ _{RL}\mathcal{I}_{t}^{\beta}\nabla{c}
\end{equation}
This condition ensures both conservation of mass and the proper units of flux at the boundary: [conc $\times$ L\textsuperscript{2}/T]. 

\section*{Stretched-Time Fractional Diffusion}
\begin{figure}
    \centering
    \includegraphics[scale = 0.6]{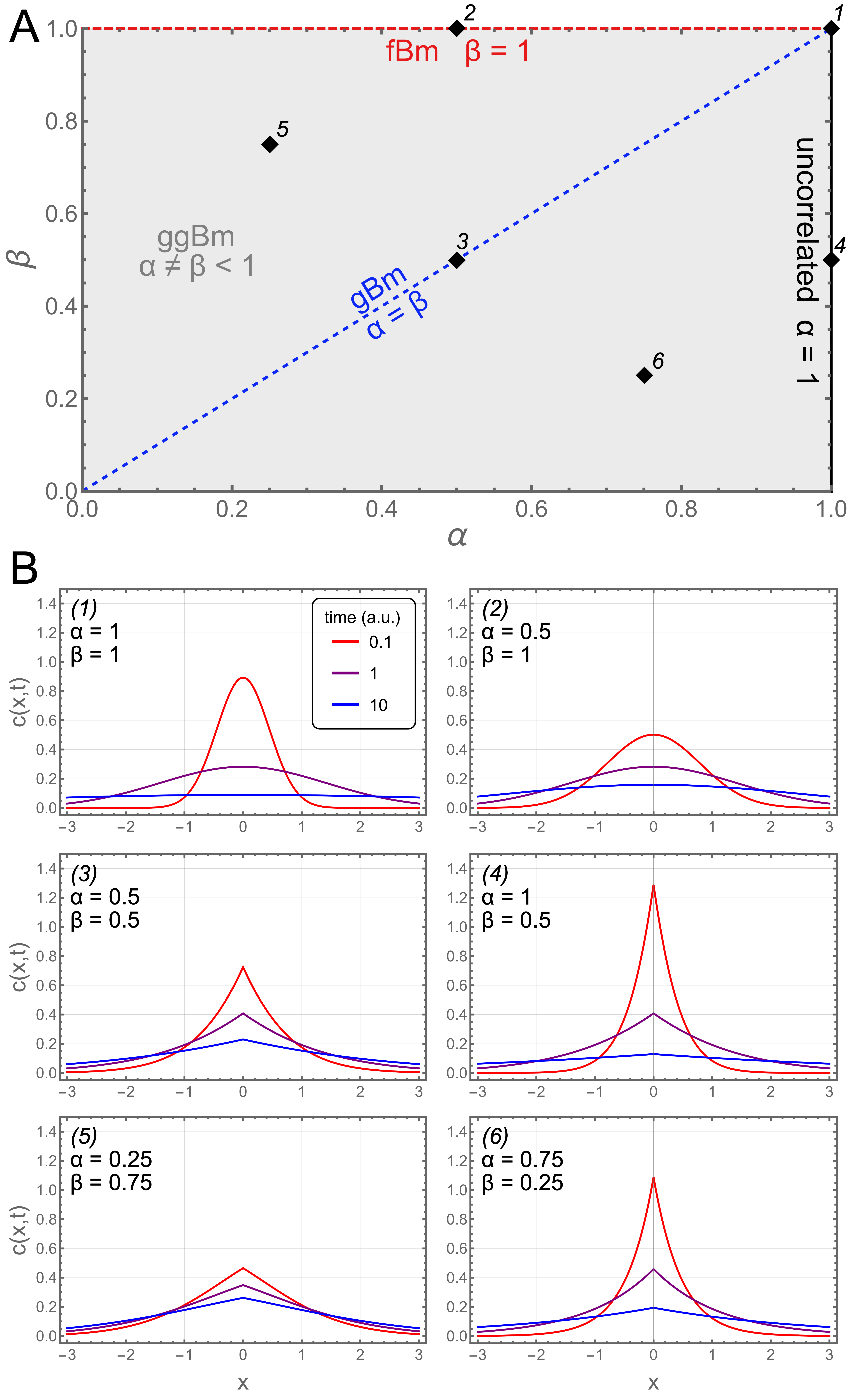}
    \caption{(A) Parameter space of $\alpha$ and $\beta$ showing associated stochastic processes. (B) Plotted results (Eqn. \ref{eqn:STFMWright}) (equivalent to single particle position PDF)  with $D_{\alpha\beta}=1$ for various combinations of $\alpha$ and $\beta$ (marked in A).}
    \label{fig:ggBm}
\end{figure}

\begin{figure}
    \centering
    \includegraphics[scale = 0.6]{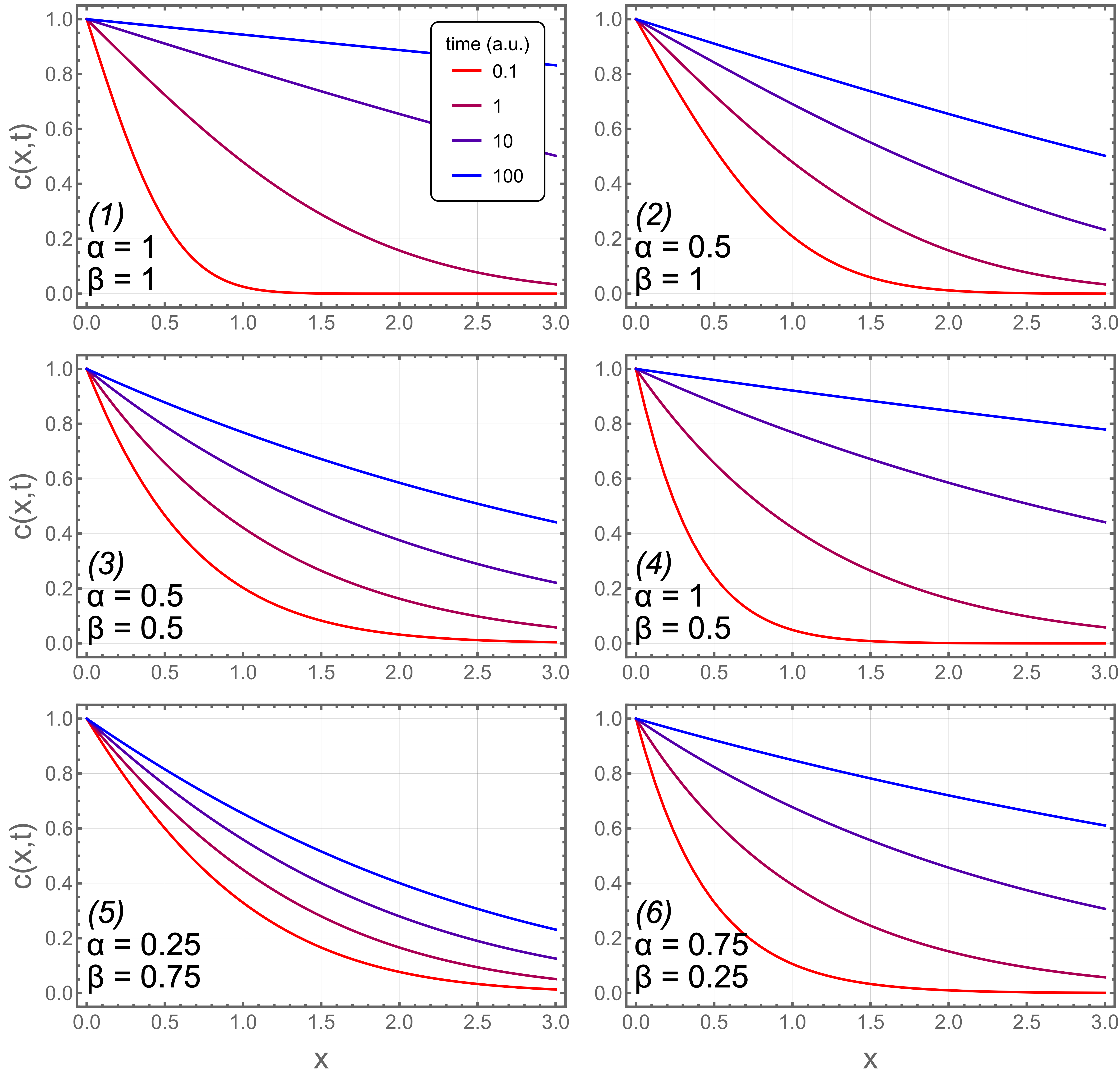}
    \caption{Plotted results (Eqn. \ref{eqn:STFKErr}) with $D_{\alpha\beta}=1$ for various combinations of $\alpha$ and $\beta$ (marked in Fig. \ref{fig:ggBm})A}
    \label{fig:ggBmK}
\end{figure}

We have considered ST and TF diffusion as two distinct systems, but these can be further generalized through a single stretched-time fractional (STF) diffusion equation for values $\alpha\in(0,2)$ and $\beta\in(0,1]$:
\begin{equation*}
    \frac{\partial c}{\partial t} = \frac{\alpha}{\beta}t^{\frac{\alpha}{\beta}-1}D_{\alpha\beta} \ _{RL}\mathcal{D}_{t^{\alpha/\beta}}^{1-\beta}\frac{\partial^{2}c}{\partial x^{2}}
\end{equation*}
Here, if $\alpha=\beta<1$, this reduces to TF diffusion, while if $\beta=1$ and $\alpha\neq1$, this reduces to ST diffusion, and if $\alpha=\beta=1$, this reduces to Fickian diffusion. The Green's function for the Cauchy problem given this equation is
\begin{equation}
    G_{\alpha,\beta}(x,t) = \frac{1}{\sqrt{4D_{\alpha\beta}t^{\alpha}}}M_{\beta/2}\left(\frac{|x|}{\sqrt{D_{\alpha\beta}t^{\alpha}}}\right)
    \label{eqn:STFMWright}
\end{equation}
which is similar in form to the Green's function for the TF diffusion equation (\ref{eqn:1DTFInfiniteGreensFunc}), suggesting that the results derived in Section V can be easily translated into STF diffusion problems. For instance, the solution to the so-called signaling problem (\S V.B.1) for the STF diffusion equation is
\begin{equation}
    c(x,t) = c_{0}K_{\beta/2}\left(\frac{x}{\sqrt{D_{\alpha\beta}t^{\alpha}}}\right)
    \label{eqn:STFKErr}
\end{equation}\\
\indent Notably, the STF diffusion formulation corresponds to a stochastic process generated by generalized gray Brownian motion (ggBm). Futher, the special cases of TF and ST diffusion correspond to the special cases of ggBm: gray Brownian motion (gBm) and fractional Brownian motion (fBm). In the context of mass transport, the dynamics of the MSD of a single particle, or distribution, is equally described by only the parameter $\alpha$: $\text{MSD}\sim t^{\alpha}$. This parameter also defines the self-similarity in the time-evolution of this system -- the Hurst exponent which characterizes fBm self-similarity ($B_{H}(ct) = c^{H}B_{H}(t)$ for some characteristic step-size probability density function $B$) is indeed $H = \alpha/2$. Small values of $\alpha$ (\textit{i.e.} $\alpha<1$) correspond to nonlinear temporal scaling, which can arise from various physical phenomena (trapping, crowding, etc.) in a medium. On the other hand, $\beta$ describes the shape, in particular, the deviation from a Gaussian shape, of either the probability density function (PDF) of a single particle, or the concentration. Small values of $\beta$ (\textit{i.e.} $\beta<1$) correspond to M-Wright functions with sharp peaks near zero and relatively fat tails. Such a PDF for a single particle suggests a heterogeneous medium with local effects, or a highly disordered medium. Plots of (Eqn. \ref{eqn:STFMWright}) and (Eqn. \ref{eqn:STFKErr}) for varying ggBm parameters are shown in Figures \ref{fig:ggBm}-\ref{fig:ggBmK}. We may note the diverse mass transport behavior for these systems -- most interesting is the behavior of (Eqn. \ref{eqn:STFKErr}): the effect of decreasing $\alpha$ limits chemical penetration through a material to a greater degree than decreasing $\beta$. That is, the physical effect of ST diffusion will play a greater role in mitigating mass transport in such a system than TF diffusion. Given the correlation between ggBm and STF diffusion, one can probe the physical nature of mass transport of a molecular species through a medium at a broad level (\textit{e.g.} is there crowding? is the material homogeneous?) through observing the collective diffusion of small molecules in the material (as we do in Part II).

\begin{landscape}
\begin{table}
\small
\centering
\caption{\label{tbl:Translations} Equivalent mathematical structures for systems defined through master diffusion equations.}
\begin{tabular}{lcccc} 
\hline
     & Fickian & ST & TF & STF \\
    \hline
    \hline
    Kernel/Gaussian & $\frac{1}{\sqrt{4\pi Dt}}\exp{\left(-\frac{x^{2}}{4Dt}\right)}$ & $\frac{1}{\sqrt{4\pi D_{\alpha}t^{\alpha}}}\exp{\left(-\frac{x^{2}}{4D_{\alpha}t^{\alpha}}\right)}$ & $\frac{1}{\sqrt{4D_{\beta}t^{\beta}}}M_{\beta/2}\left(\frac{|x|}{\sqrt{D_{\beta}t^{\beta}}}\right)$ & $\frac{1}{\sqrt{4D_{\alpha\beta}t^{\alpha}}}M_{\beta/2}\left(\frac{|x|}{\sqrt{D_{\alpha\beta}t^{\alpha}}}\right)$ \\
    Error Function & $\text{erf}{\left(\frac{x}{\sqrt{4Dt}}\right)}$ & $\text{erf}{\left(\frac{x}{\sqrt{4D_{\alpha}t^{\alpha}}}\right)}$ & $N_{\beta/2}\left(\frac{x}{\sqrt{D_{\beta}t^{\beta}}}\right)$ & $N_{\beta/2}\left(\frac{x}{\sqrt{D_{\alpha\beta}t^{\alpha}}}\right)$ \\
    Comp. Error Function & $\text{erfc}{\left(\frac{x}{\sqrt{4Dt}}\right)}$ & $\text{erfc}{\left(\frac{x}{\sqrt{4D_{\alpha}t^{\alpha}}}\right)}$ & $K_{\beta/2}\left(\frac{x}{\sqrt{D_{\beta}t^{\beta}}}\right)$ & $K_{\beta/2}\left(\frac{x}{\sqrt{D_{\alpha\beta}t^{\alpha}}}\right)$\\
    Temporal Propagator & $\exp{\left(-\lambda^{2}Dt\right)}$ & $\exp{\left(-\lambda^{2}D_{\alpha}t^{\alpha}\right)}$ & $E_{\beta}\left(-\lambda^{2}D_{\beta}t^{\beta}\right)$ & $E_{\beta}\left(-\lambda^{2}D_{\alpha\beta}t^{\alpha}\right)$
\end{tabular}
\end{table}
\end{landscape}

\section*{Conclusion}
In this work, we have shown how one may develop solutions to either the ST, TF, or STF diffusion equations, the latter two through the use of two well-known special functions, the Mittag-Leffler and M-Wright functions, as well as two newly developed special functions, the fractional error and complementary error functions. With these, one may use appropriate solutions developed for the standard, Fickian diffusion equation and translate them into the their equivalent formulations to describe a system with anomalous diffusion (summarized in Table \ref{tbl:Translations}). The use of these frameworks can inform a researcher as to the mass transport properties of a medium governing single particle Brownian motion from the collective behavior of a diffusing concentration profile. Looking beyond these problems, one may consider the same framework of functional replacement for other systems with time fractional derivatives, such as the time fractional wave equation, or time fractional Schrodinger equation.

\section*{Conflicts of interest}
There are no conflicts of interest to declare.

\section*{Acknowledgements}
This publication was supported by U.S. Environmental Protection Agency (EPA) STAR Center
Grant \#84003101. Its contents are solely the responsibility of
the grantee and do not necessarily represent the official views of the U.S. EPA. Further, U.S. EPA does not endorse the purchase of any commercial products or services mentioned in
the publication.

\section*{Author Contributions}
Conceptualization, N.G.H.; methodology, N.G.H.; software, N.G.H.; validation, N.G.H.; formal analysis, N.G.H; investigation, N.G.H.; resources, \textit{n/a}; data curation, \textit{n/a}; writing---original draft preparation, N.G.H.; writing---review and editing, N.G.H. and M.S.H.; visualization, N.G.H.; supervision, M.S.H.; project administration, M.S.H.; funding acquisition, M.S.H. All authors have read and agreed to the published version of the manuscript.

\appendix

\section*{Appendix}

\section{Notation}
Throughout, we use three integral transforms: the Laplace, Fourier, and Mellin transforms. We define the Laplace transform, $\mathcal{L}:t\rightarrow s$, of a well-behaved function $f(t)$ as
\begin{align}
    \tilde{f}(s) &= \mathcal{L}[f(t)] = \int_{0}^{\infty}e^{-st}f(t)dt \\
    f(t) &= \mathcal{L}^{-1}[\tilde{f}(s)] = \int_{Br}e^{st}\tilde{f}(s)ds
\end{align}
where $Br$ denotes the Bromwich path in the complex plane. We define the Fourier transform, $\mathcal{F}:x\rightarrow k$, of a well-behaved function $f(x)$ as
\begin{align}
    \hat{f}(k) &= \mathcal{F}[f(x)] = \int_{-\infty}^{\infty}e^{ikx}f(x)dx \\
    f(x) &= \mathcal{F}^{-1}[f(k)] = \frac{1}{2\pi}\int_{-\infty}^{\infty}e^{-ikx}\hat{f}(k)dk
\end{align}
We define the Mellin transform, $\mathcal{M}:r\rightarrow s$, of a well-behaved function $f(r)$ as
\begin{align}
    f^{*}(s) &= \mathcal{M}[f(r)] = \int_{0}^{\infty}r^{s-1}f(r)dr \\
    f(r) &= \mathcal{M}[f^{*}(s)] = \frac{1}{2\pi i}\int_{Br}r^{-s}f^{*}(s)ds
\end{align}

\section{Time Fractional Superdiffusion}
\begin{figure}
    \centering
    \includegraphics[scale = 1]{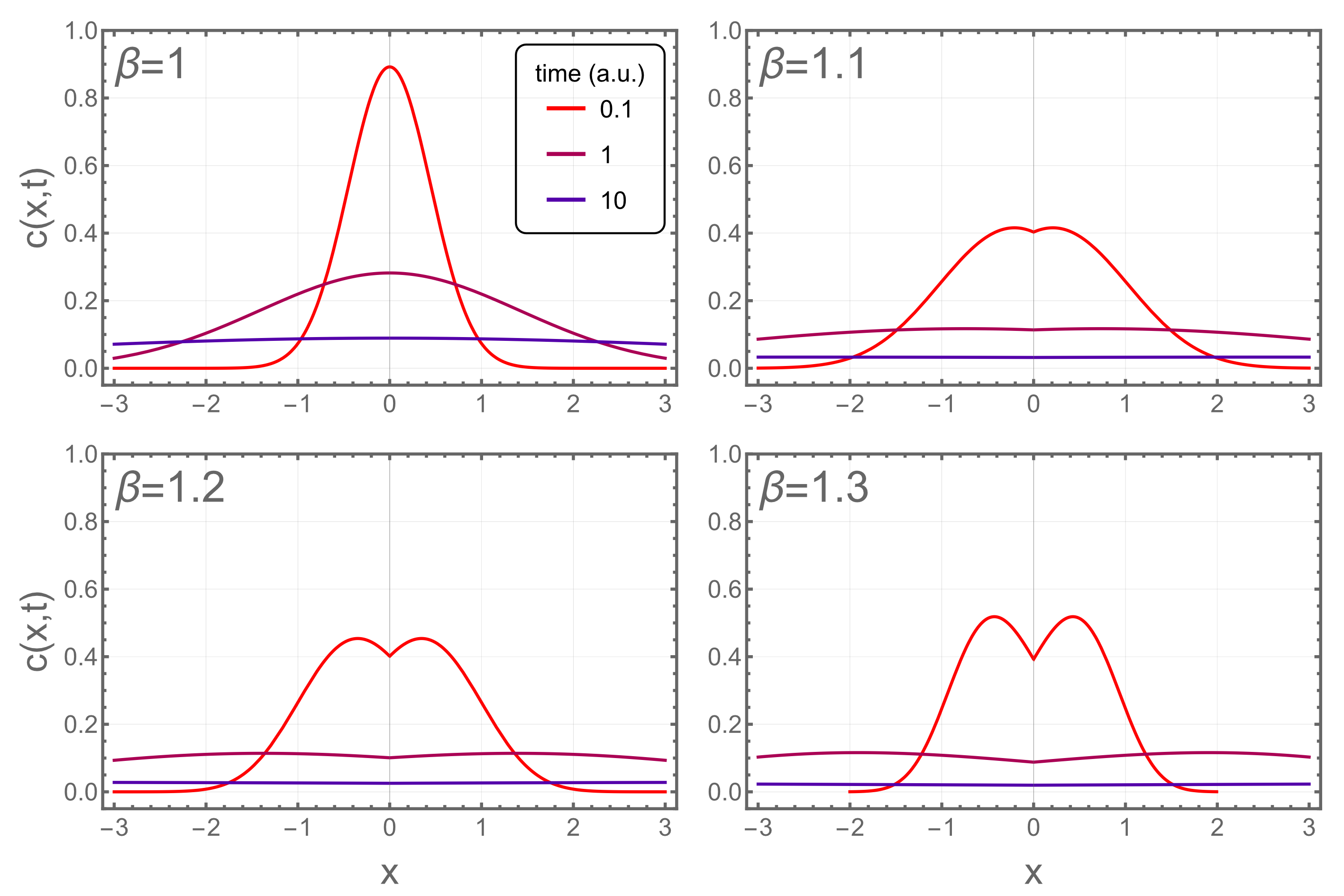}
    \caption{TF diffusion in an infinite domain with initial condition $c(x,t) = \delta(x)$. Plotted results of Eqn. \ref{eqn:TFSuperdiffusion} with $D_{\beta}=1$ for superdiffusive values of $\beta$.}
    \label{fig:AppendixInfDomainDeltaICSuper}
\end{figure}

\begin{figure}
    \centering
    \includegraphics[scale = 0.6]{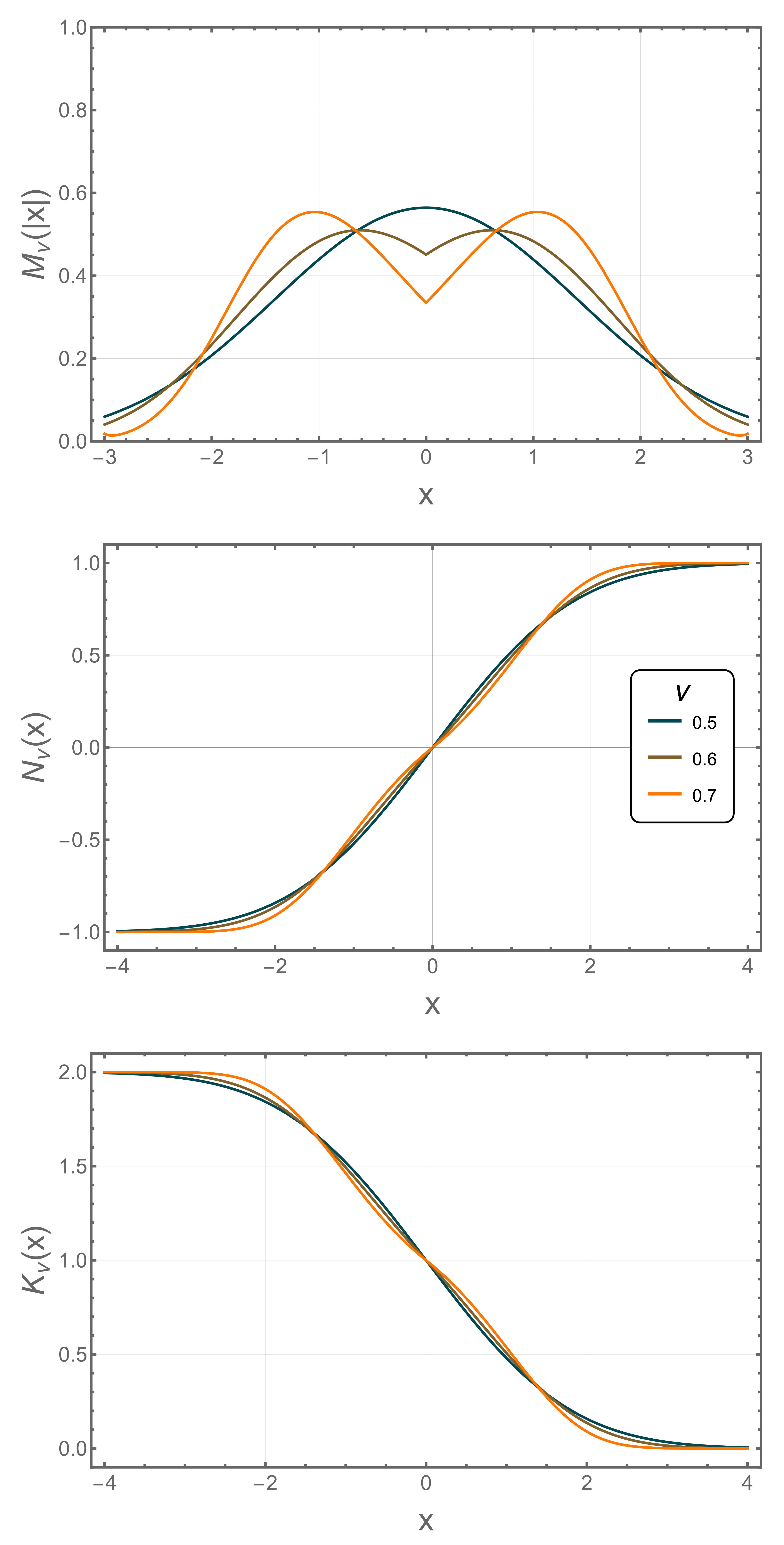}
    \caption{Plots of the M-Wright (top), fractional error (middle), and fractional complementary error functions (bottom) over superdiffusive TF derivative order $\nu$, where $\nu = 0.5$ recovers Fickian diffusion.}
    \label{fig:AppendixFunctionPlots}
\end{figure}

Here, let us consider the superdiffusive case ($1<\beta<2$) and show how the framework of TF diffusion interpolates between the diffusion equation and wave equation. Let us consider first the Cauchy problem, where $c(x,0) = \delta(x)$, $c(x\rightarrow\pm\infty) = 0$, and $x\in\mathbb{R}$. First, applying the Fourier transform,
\begin{equation}
    \frac{\partial^{\beta}\hat{c}}{\partial t^{\beta}} = -D_{\beta}k^{2}\hat{c}(k,t)
\end{equation}
with initial condition $\hat{c}(k,0) = 1$. Then, applying the Laplace transform,
\begin{equation}
    s^{\beta}\tilde{\hat{c}}(k,s) -s^{\beta-1}\hat{c}(k,0) - s^{\beta-2}\frac{\partial\hat{c}}{\partial t}\bigg\rvert_{t=0} = -D_{\beta}k^{2}\tilde{\hat{c}}(k,s)
\end{equation}
Applying initial conditions, this reduces to
\begin{equation*}
     s^{\beta}\tilde{\hat{c}}(k,s) -s^{\beta-1} = -D_{\beta}k^{2}\tilde{\hat{c}}(k,s)
\end{equation*}
or, rearranging,
\begin{equation}
    \tilde{\hat{c}}(k,s) = \frac{s^{\beta-1}}{s^{\beta} + D_{\beta}k^{2}}
\end{equation}
This is in fact identical to the subdiffusive solution
\begin{equation}
    c(x,t) = \frac{1}{\sqrt{4D_{\beta}t^{\beta}}} M_{\beta/2}\left(\frac{|x|}{\sqrt{D_{\beta}t^{\beta}}}\right)
    \label{eqn:TFSuperdiffusion}
\end{equation}
Note the limiting case when $\beta = 2$, at which we recover the solution \cite{Mainardi2010aF}
\begin{equation}
    c(x,t) = \frac{1}{2}\left[\delta\left(x-\sqrt{D_{2}}t\right) + \delta\left(x+\sqrt{D_{2}}t\right)\right]
\end{equation}
which is of course the solution to the wave equation for wave speed $\sqrt{D_{2}}$. In this sense, the TF diffusion equation smoothly interpolates between diffusion and wave propagation. Plots of the solution given by Eqn. \ref{eqn:TFSuperdiffusion} are given in Figure \ref{fig:AppendixInfDomainDeltaICSuper}. \\
\indent In general, solutions developed for subdiffusion will hold in the superdiffusive case, as the M-Wright function $M_{\nu}$ is well defined for $\nu\in[0,1]$. In practice, finding numerical values for the M-Wright function and associated functions ($N$ and $K$) is difficult, as the infinite sum in the series form of these functions tends to infinity at sufficiently large arguments for any arbitrary termination of the series. Plots of the M-Wright function, fractional error function, and fractional complementary error function for superdiffusive systems are shown in Figure \ref{fig:AppendixFunctionPlots} for relatively small orders $\nu\leq 1.3$, below which we can approximate these functions well. 

\section{Space Fractional Diffusion}
\begin{figure}
    \centering
    \includegraphics[scale = 1.25]{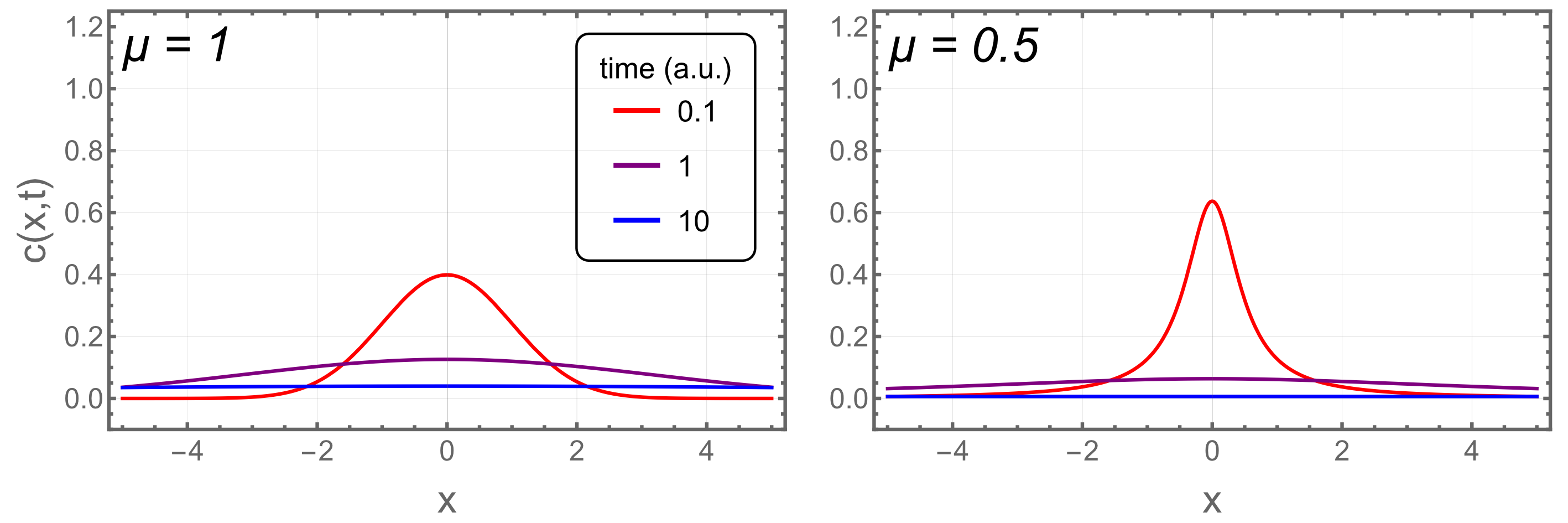}
    \caption{Plots of Eqn. \ref{eqn:SpaceFra} with $D_{\mu} = 5$ for (left) $\mu = 1$ and (right) $\mu = 0.5$.}
    \label{fig:AppendixLevyStable}
\end{figure}

Consider a space fractional diffusion equation given as:
\begin{equation}
    \frac{\partial c}{\partial t} = D_{\mu}\frac{\partial^{2\mu}c}{\partial x^{2\mu}}
\end{equation}
where the space-fractional derivative is defined as a Riesz fractional derivative, defined through its essential property:
\begin{equation*}
    \mathcal{F}\left[\frac{\partial^{\mu}f}{\partial x^{\mu}}\right] = -|k|^{\mu}\hat{f}(k)
\end{equation*}
The structure of the Riesz derivative and its properties have been well-studied particularly in the context of a fractional Schrodinger equation \cite{Ding2014,Prado2015,Bayin2016,Tarasov2023}. Let us solve the Cauchy problem for this master equation where $x\in\mathbb{R}$, $t>0$, and with have initial condition $c(t=0) = \delta(x)$ (that is, solving for the Green's equation). Taking the Laplace transform, we find
\begin{equation}
    s\tilde{c}-\delta(x) = D_{\mu}\frac{\partial^{2\mu}c}{\partial x^{2\mu}}
\end{equation}
Then, taking a Fourier transform, we find
\begin{equation}
    s\hat{\tilde{c}} - 1 = -D_{\mu}|k|^{2\mu}\hat{\tilde{c}}
\end{equation}
Therefore,
\begin{equation}
    \hat{\tilde{c}} = \frac{1}{s+D_{\mu}|k|^{2\mu}}
\end{equation}
Then, taking the inverse Laplace transform,
\begin{equation}
    \hat{c} = \exp{(-D_{\mu}|k|^{2\mu}t)}
\end{equation}
This is in fact the characteristic function corresponding to a class of functions -- the Lévy stable distributions \cite{Mainardi2010aF}. These are defined via their characteristic function (\textit{i.e.} their Fourier transform):
\begin{equation}
    \mathcal{F}\left[L_{\mu}^{\theta}(x)\right] = \exp{\left(|k|^{\mu}e^{i\ \text{sgn}(k)\theta\pi/2}\right)}
\end{equation}
Then, taking $\theta=0$ (corresponding to zero skewness), and without loss of generality, one may define
\begin{equation}
    \mathcal{F}\left[\frac{1}{t^{\mu}}L_{\mu}^{0}\left(\frac{x}{t^{\mu}}\right)\right] = \exp{(-t|k|^{\mu})}
\end{equation}
So, it follows that the solution to the equation with given conditions is
\begin{equation}
    c(x,t) = \frac{1}{(D_{\mu}t)^{1/2\mu}}L_{2\mu}^{0}\left(\frac{x}{(D_{\mu}t)^{1/2\mu}}\right)
    \label{eqn:SpaceFra}
\end{equation}
Plots of this solution for both $\mu = 1$ and $\mu = 0.5$ are shown in Figure \ref{fig:AppendixLevyStable}. We recover qualitatively similar behavior as for time fractional diffusion in so much as the distribution is tightened. At a smaller value of $\mu$, the concentration is spread further more quickly than in the case of purely Fickian diffusion. Notably, when $\mu = 1$, this recovers the standard Gaussian solution for Fickian diffusion, given that
\begin{equation*}
    L_{2}^{0}(x) = \frac{1}{\sqrt{4\pi}}\exp{\left(-\frac{x^{2}}{4}\right)}
\end{equation*}
Moreover, this class of functions is closely related to the M-Wright function through the definitions:
\begin{align}
    L_{\mu}^{-\mu}(x) &= \frac{\mu}{x^{\mu+1}}M_{\mu}(x^{-\mu}) \\
    L_{\mu}^{2-\mu}(x) &= \frac{1}{\mu}M_{1/\mu}(x)
\end{align}

\bibliographystyle{ieeetr}
\bibliography{citations}

\end{document}